\documentclass[11pt]{article}

\pdfoutput=1
\usepackage{amsfonts}
\usepackage{amsmath}
\usepackage{amssymb}
\usepackage{amsthm}
\usepackage[title]{appendix}
\usepackage{authblk}
\usepackage[english]{babel}
\usepackage{cancel}
\usepackage{cite}
\usepackage{color}
\usepackage{colortbl}
\usepackage{comment}
\usepackage{enumerate}
\usepackage{enumerate}
\usepackage{float}
\usepackage[T1]{fontenc}
\usepackage{framed}
\usepackage{fullpage}
\usepackage[final]{graphicx}
\usepackage[final]{hyperref}
\usepackage[utf8]{inputenc}
\usepackage{latexsym}
\usepackage[modulo]{lineno}
\usepackage{lipsum}
\usepackage{mathrsfs}
\usepackage{paralist}
\usepackage{tikz}

\hypersetup{
  colorlinks   = true, 
  urlcolor     = blue, 
  linkcolor    = {blue!55!black}, 
  citecolor   = red 
}

\advance \topmargin by -\headheight
\advance \topmargin by -\headsep
\evensidemargin \oddsidemargin
\usepackage{ifdraft}
\ifdraft{\usepackage{showlabels}}{}
\ifdraft{\linenumbers}{}
\usepackage[refpage]{nomencl}

\newcommand{\cC}{\mathcal{C}}
\newcommand{\cD}{\mathcal{D}}

\newcommand{\cF}{\mathcal{F}}

\newcommand{\cI}{\mathcal{I}}

\newcommand{\cL}{\mathcal{L}}
\newcommand{\cM}{\mathcal{M}}

\newcommand{\cP}{\mathcal{P}}
\newcommand{\cQ}{\mathcal{Q}}

\newcommand{\cT}{\mathscr{T}}

\newcommand{\cV}{\mathcal{V}}
\newcommand{\cW}{\mathcal{W}}

\newcommand{\bC}{\mathbb{C}}

\newcommand{\bE}{\mathbb{E}}

\newcommand{\bN}{\mathbb{N}}

\newcommand{\bT}{\mathbb{T}}

\newcommand{\abs}[1]{\left| #1 \right|}

\newcommand{\bra}[1]{\langle #1|}

\newcommand{\ham}{\mathcal{H}}

\newcommand{\innerprod}[2]{\langle #1 | #2 \rangle}

\newcommand{\ket}[1]{\left\vert #1 \right\rangle}

\newcommand{\Mk}{\cM_{k\times k}}

\newcommand{\norm}[1]{\left\| #1 \right\|}

\newcommand{\parent}[1]{\left(#1\right)}
\newcommand{\poly}[1]{\mathrm{poly(} #1 \mathrm{)}}

\newcommand{\prob}[1]{\Pr\left[ #1 \right]}
\newcommand{\probover}[2]{\Pr_{#1}[ #2 ]}

\newcommand{\tr}[1]{tr\left( #1 \right)}

\newcommand{\BPP}{\mathsf{BPP}}
\newcommand{\BQP}{\mathsf{BQP}}

\newcommand{\ILSCC}{\textit{ILSCC}}
\newcommand{\IP}{\mathsf{IP}}

\newcommand{\MA}[1]{\mathsf{MA\left[ #1 \right]}}

\newcommand{\AM}{\mathsf{AM}}

\newcommand{\p}{\mathsf{P}}

\newcommand{\PostBQP}{\mathsf{PostBQP}}

\newcommand{\PP}{\mathsf{PP}}

\newcommand{\PsP}{\p^{\#\p}}

\newcommand{\PSPACE}{\mathsf{PSPACE}}

\newcommand{\sP}{\#\p}

\theoremstyle{plain}

\newtheorem{theorem*}{Theorem}

\newtheorem{claim}{Claim}

\theoremstyle{definition}
\newtheorem{defn}{Definition}[section]

\theoremstyle{remark}

\DeclareMathOperator*{\argmax}{\mathsf{argmax}}
\DeclareMathOperator*{\E}{\mathbb{E}}

\newcommand{\rmarker}[1]{{\ifdraft{\color{red}#1}{#1}}}
\newcommand{\bmarker}[1]{{\ifdraft{\color{blue}#1}{#1}}}

\newcommand{\remove}[1]{}

\begin{document}

\title{On Information-Theoretic Classical Verification of Quantum Computers}

\author{Ayal Green} 
\affil{School of Computer Science and Engineering, The Hebrew University, Jerusalem 91904,  Israel}
\maketitle

\abstract{Quantum inspired protocols e.g. \cite{AAV13, AG17} attempt to achieve a single-prover interactive protocol where a classical machine can verify quantum computations in an information-theoretically secure manner. 
We define a family of protocols which seem natural for verifying quantum computations and generalizes such known protocols, namely those of \cite{AAV13, AG17}. We show that any protocol from this family is bound to require an extremely powerful prover, much like the classical protocols of \cite{LFKN92} and \cite{Sha92}. Using our analysis, we also hint at possible ways one might try to realize a protocol where the prover can be weaker, namely a quantum computer (i.e. a $\BQP$ machine). 
}

\section{Introduction}  \label{sec: intro}

One of the main open questions in Quantum Complexity Theory has to do with delegation of quantum computations. More specifically, whether a $\BPP$ verifier can delegate polynomial time quantum computations (namely computations in $\BQP$) to a $\BQP$ prover which can not be trusted. 
The initial protocols for the problem \cite{ABOE10, BFK09} (with follow up work \cite{FM16, ABOEM17, FK17}) required the $\BPP$ verifier to process constantly many quantum bits. Alternatively, \cite{RUV13} considered a setting where a strictly classical verifier is given access to multiple, entangled, non-interacting provers. However, despite remarkable advancement in this line of work over the past decade (see e.g. \cite{BH13, FH13, FV15, Ji16a, Ji17, NV17, CGJV19} and notably \cite{NV18, JNVWY20}), the basic requirements on the provers were not relaxed. Revisitng the original setting where a strictly classical verifier interacts with a single $\BQP$ prover, another notable result by \cite{Mah18} showed a protocol which achieves the goal, but where soundness relies on computational assumptions. It is still open whether information-theoretically sound verification (where security is required against dishonest provers with unbounded computational power)
can be done with a single prover, when the verifier is entirely classical, namely a $\BPP$ machine.  

On the other hand, we do know that strictly classical interactive protocols are strong enough for information-theoretic verification of $\BQP$ computations, if the prover is sufficiently strong. The celebrated $\PsP \subseteq \IP$ and $\IP=\PSPACE$ 
results \cite{LFKN92,Sha92} show this, as $\BQP \subseteq \PP$ \cite{BV93}. Alas, attempting to adapt the protocols of \cite{LFKN92,Sha92}, namely the \textit{sum-check} protocol, to be used with a weaker $\BQP$ prover seems to be problematic. In order to prove a $\BQP$ problem using the \textit{sum-check} protocol, one must first reduce it to a general $\sP$ problem - which is no longer solvable by a $\BQP$ machine. To address this issue, \cite{AG17} suggest an alternative protocol, which is more natural to the quantum setting. In \cite{AG17} the problem is formulated in terms of local quantum operations and measurements, and they show that a $\PostBQP$ \cite{Aar05} prover, which may evaluate quantum values to within inverse-exponential precision, suffices for it. Unfortunately, \cite{AG17} also show that if their protocol is relaxed by allowing inverse-polynomial errors, such that an honest $\BQP$ prover may pass it, the modified protocol is no longer secure and there exists a cheating strategy.

\paragraph{Our contribution.}  \label{sec: contribution}

We present an in depth discussion of the problems that arise when trying to devise protocols for verifying quantum computations using a single $\BQP$ prover. To do so, we define a family of protocols, called \textit{Inexact Linear Scalar Consistency Checking (ILSCC)} protocols - and prove strong limitations on them. In $\ILSCC$ protocols, for each round $i$, there is a scalar $u_i$ which is evaluated at the beginning of the round, and a scalar $v_i$ which is evaluated at the end of the round. A key characteristic of $\ILSCC$ protocols is that the verification process only consists of comparing $u_i$ and $v_{i-1}$ at each round. Another key characteristic of these protocols is that these values are computed by linear functionals. Moreover, we consider such protocols where the comparisons are performed up to an inverse polynomial accuracy (see Section \ref{sec: LSCCs} for exact definition). These are natural protocols for verifying $\BQP$ computations, as according to the circuit model all that a $\BQP$ machine does are linear manipulations of vector states, and as $\BQP$ machines can only provide an inverse-polynomially accurate estimation of their inner state, due to statistical errors. In particular, the $\ILSCC$ family of protocols generalizes the protocols of \cite{AAV13, AG17}. For an in-depth discussion of the generality of $\ILSCC$ protocols, see Section \ref{sec: open questions}. We analyze $\ILSCC$ protocols based on a parameter which we call the \textit{expected next value}. This value has to do with the maximal expected decrease of the consistency scalar $v_i$ between one round to the following\footnote{For a complete explanation of what the maximum and expectation are taken over, see Section \ref{sec: ILSCC limitations}.}. Based on this expected decrease we show the following (see Section \ref{sec: ILSCC limitations} for the exact statements and proofs): \\

\textbf{Theorem \ref{cl:decreasing max value}} (Informal): There is a cheating strategy against all \textit{ILSCC} protocols where the scalar consistency value is expected to decrease by a multiplicative constant factor.
\\

\textit{Proof overview:} Theorem \ref{cl:decreasing max value} follows from showing that if the scalar consistency value is expected to decrease by a constant factor in each round - then a cheating prover can send erroneous values in each round $i$ such that the error $\delta_i$, which is the difference between the actual evaluated $v_i$ and the value $v'_i$ which \textit{would have been evaluated} in this round had the prover been honest, will also reduce by a constant factor. This error reduction in every round adds up to an exponential reduction at the end of the protocol. But such errors are not caught by the verifier, which allows errors of up to an inverse polynomial. \\

\textbf{Theorem \ref{cl:non-decreasing max value}} (Informal): There is no interactive proof for $\BQP$ computations of the form of \textit{ILSCC}, where the scalar consistency value doesn't decrease (i.e. stays the same\footnote{The exact definitions are such that an \textit{increase} is not possible.}), unless $\BQP = \BPP$.
\\

\textit{Proof overview:} Theorem \ref{cl:non-decreasing max value} follows from the following reasoning: if the evaluated values do not change between rounds, then these rounds are essentially redundant. This means that the intermediate rounds can be removed, resulting with a protocol where there are only 2 rounds of communication between the verifier and the prover. However, as the protocol is only accurate up to an inverse-polynomial, these rounds only include logarithmically many bits of information sent from the prover to the verifier. Therefore, the prover can iterate over all possible values he expects from the prover, and by that simulate the protocol on his own. This means that, had there existed such a protocol to verify $\BQP$ computations, this protocol could also be used to solve $\BQP$ in $\BPP$. \\ 

Our analysis highlights inherent problems in devising protocols for verifying $\BQP$ computations. We hope this steers future research towards approaches which may be viable for verifying quantum computers. In particular, our analysis brings forth open questions regarding cases where it does not apply. Notably, it still leaves some window of values for the \textit{expected next value} in which a secure \textit{ILSCC} protocol might exists. This is discussed at length in Section \ref{sec: open questions}.

\paragraph{Related work.} As we had already mentioned, our $\ILSCC$ family of protocols generalizes the protocols of \cite{AAV13, AG17}. It is also natural to consider the relation between $\ILSCC$ and known classical protocols, namely \cite{LFKN92, Sha92}. At first glance, it might seem like the protocol of \cite{LFKN92} is indeed an $\ILSCC$ protocol. The protocol consists of the prover sending a polynomial at each round, and the verifier comparing an evaluation of the polynomial from round $i$ with a linear combination of evaluations of the polynomial from round $i+1$. These are comparisons of scalar values which result from linear operations. However, a closer examination of the $\ILSCC$ definition in Section \ref{sec: LSCCs} also constrains the operator norm of an associated linear operator, which is undefined when working in a finite field. The protocol of \cite{Sha92} shares this dissimilarity. In addition, the consistency checks of \cite{Sha92} also include multiplying different evaluations of a polynomial - which is not a linear operation. 

\paragraph{Paper organization.}
General preliminaries and notations are provided in Section \ref{sec: background}. We define a family of generalized protocols, namely $\ILSCC$, and associated notions in Section \ref{sec: LSCCs}. We analyse the inherent difficulty of using $\ILSCC$ protocols with a $\BQP$ prover in Section \ref{sec: ILSCC limitations}. Finally, we discuss the generality of $\ILSCC$ protocols as well as open questions regarding possible approaches to devise verification protocols with a $\BQP$ prover in Section \ref{sec: open questions}.

\section{Background} \label{sec: background} 

\subsection{Notations}
We use the following notations throughout the paper:

\begin{itemize}

	\item For a local gate $g_i$, we define $G_i=g_i\otimes I$ to be its $n$ qubits expansion (where $n$ is the number of qubits in the entire system).
		
	\item Similarly, for local unitaries $u_i$ or $h_i$ we use $U_i$ and $H_i$ (respectively) to denote their n qubits expansions.

	\item Let $\rho$ be an operator on hilbert space $\ham = \ham_A \otimes \ham_B$, and let $z$ be an operator on $\ham_A$. Given an orthonormal basis $\{\ket{i}\}_i$ for $\ham_A$ and $\{\ket{l}\}_l$ for $\ham_B$, such that we can write 
$\rho =\sum_{i,k,j,l}\rho_{i,j,k,l}\ket{i}\bra{j}\otimes \ket{k}\bra{l}$, we define the reduction of $\rho$ to the qubits on which $z$ operates as
\[
\rho|_z=tr_{\ham_B}(\rho)=
tr_{\ham_B}\left(
\sum_{i,j,k,l}{\rho_{i,k,j,l}\ket{i}\bra{j}\otimes\ket{k}\bra{l}}\right)=\sum_{i,j}
{\sum_k\rho_{i,j,k,k}}\ket{i}\bra{j}\
\]
    \item We use the notation $\approx_\mu$ to denote equality to within $\mu$ precision. That is, for scalars $a,b$:
    \[
    a\approx_\mu b \Longleftrightarrow \abs{a-b} \leq \mu
    \]
    \item We use $\bC$ to denote the set of complex numbers.
\end{itemize}

\subsection{BQP}

\begin{defn}[$\BQP$, due to \cite{BV93}] The complexity class $\BQP$ is the set of languages which are accepted with probability 2/3 by some polynomial time Quantum Turing Machine.
\end{defn}

\subsection{Top Row Matrix \cite{AG17}} \label{sec: top row}
A \textit{top row matrix} is defined as follows:
\begin{defn}[top row matrix] \label{def:toprow}
Given a $2^n\times2^n$ matrix $B$, we define it's \textit{top row matrix} to be: $\vert0^n\rangle \langle 0^n \vert B$, where $\vert0^n\rangle$ is the computational basis state on $n$ qubits $\vert00\cdots0\rangle$. By referring to the \textit{top row matrix} of a quantum circuit on $n$ qubits which is given as a sequence of local gates 
$g_T\ldots g_1$ we mean the \textit{top row matrix} of 
$B=G_T\cdot G_{T-1}\cdots G_1$.
\end{defn}
\cite{AG17} also show the simple fact that calculating the trace of the \textbf{top row matrix} of a matrix $B$ with precision $\pm\frac{1}{6}$, where $B$ is the operator which results from compositing an arbitrary sequence of local gates $g_T,\ldots,g_1$, such that each gate $g_i$ acts non-trivially on $3$ out of $n$ input qubits and such that $T=poly(n)$, is $\BQP$-$hard$.

\subsection{The AG protocol} \label{sec:The AG protocol}
We feel our discussion of $\ILSCC$ protocols is easier to understand in the context of a specific protocol. To this end, we will present the protocol of \cite{AG17} in which a verifier $V$ verifies the trace $C$ of a \textit{top row matrix} $A$ for a quantum computation given by a sequence of local quantum gates $g_1,\cdots ,g_T$, by interacting with a prover $P$. More formally, the protocol verifies $\tr{A}=C$ 
for $A=\vert0^n\rangle \langle 0^n \vert G_T\cdot G_{T-1}\cdots G_1$ and value $C$. \\

The protocol uses the following measure to generate random unitaries on a single qubit: 

\begin{defn}[The measure $U(2)$]\label{def:U(2)} The measure $U(2)$ on unitary operators on a single qubit is defined as choosing angles $\theta, \varphi_1, \varphi_2$ uniformly at random from $[0,2\pi]$, and pick the resulting unitary $u$ to be:

\[
	u= \begin{pmatrix}   
	\cos\theta\cdot e^{i\varphi_1} & \sin\theta\cdot e^{i\varphi_2} \\
	-\sin\theta\cdot e^{-i\varphi_2} & \cos\theta\cdot e^{-i\varphi_1}
	\end{pmatrix}
\]
\end{defn}
With this measure defined, we can present the AG protocol: 
\begin{framed}
	\paragraph{The AG protocol:}
	Given a \textit{top row matrix} $A_{2^n\times 2^n}=\vert0^n\rangle \langle 0^n \vert\cdot G_T\cdot G_{T-1}\cdots G_1$, and a claim $\tr{A}=C$, the AG protocol between the verifier $V$ and prover $P$ goes as follows:
	\begin{enumerate}
		\item {\em In the 0'th round} -- $V$ asks for $M_0=A|_{g_1}$, receives back a matrix $m_0$ from $P$, and verifies that $C=\tr{m_0}$ (rejects otherwise).
 		\item {\em In the $i$'th round} -- $V$ chooses $u_i^1, u_i^2, u_i^3 \backsim U(2)$, sets $u_i = u_i^1\otimes u_i^2 \otimes u_i^3$ on the qubits on which $g_i$ operates, and asks for:
		 \[
		 M_i=(U_i\cdot U_{i-1}\cdots U_1 \cdot \vert0^n\rangle \langle 0^n \vert\cdot G_T\cdots G_{i+1})|_{g_{i+1}}
		\]
		$V$ receives back a matrix $m_i$ from $P$, and verifies that $\tr{m_i}=\tr{m_{i-1}\cdot g_i^{-1}\cdot u_i}$ (rejects otherwise).
		\item {\em After the T'th round} -- $V$ computes $\tr{M_T}=\tr{U_T\cdot U_{T-1}\cdots U_1 \cdot \vert0^n\rangle \langle 0^n \vert}$ on its own (the various $U_i$ all act on single qubits in tensor product, so this can be done efficiently) and accepts iff
		\[
		\tr{m_T}=\tr{M_T} 
		\].
	\end{enumerate}
\end{framed}

\paragraph{Completeness and soundness} \cite{AG17} show that due to the algebraic properties of the trace operator:
\begin{equation} \label{eq:consistency check}
    \tr{A}=\tr{M_0},\quad  \tr{M_i}=\tr{M_{i-1}\cdot g_i^{-1}\cdot u_i}
\end{equation}
And so their protocol has completeness 1. They also show that the protocol's soundness is 0 (see Section 5 in \cite{AG17}). In order to prove the soundness parameter, \cite{AG17} present the error matrix $\Delta_i$ for each round $i$, which is defined to be:
\begin{equation} \label{eq:AG delta}
\Delta_i = M_i-m_i
\end{equation}
And by Equations \ref{eq:consistency check}, \ref{eq:AG delta}, along with the linearity of the trace operator, they get:
\begin{equation} \label{eq:delta equivalence}
\tr{m_i}=\tr{m_{i-1}\cdot g_i^{-1}\cdot u_i} \Longleftrightarrow \tr{\Delta_i}=\tr{\Delta_{i-1}\cdot g_i^{-1}\cdot u_i}
\end{equation}
So, in order to pass round $i$'s verification it has to be the case that
\begin{equation} \label{eq:delta consistency}
\tr{\Delta_i}=\tr{\Delta_{i-1}\cdot g_i^{-1}\cdot u_i}
\end{equation}
They also show that:
\begin{equation} \label{eq:error forwarding}
    \tr{\Delta_{i-1}}\neq 0 \rightarrow \probover{u_i}{\tr{\Delta_{i-1}\cdot g_i^{-1}\cdot u_i}=0}=0
\end{equation}

In other words, if $m_i$ is different than $M_i$ then the prover has to send $m_{i+1}$ which is different than $M_{i+1}$ - otherwise he will be caught in the round's verification. This will continue until the final round, in which the verifier computes the final value on its own (as it is now a local operation) and can catch the cheating prover. 

\paragraph{Precision errors}
The AG protocol above assumed infinite precision in computations, values and probability measure used. In order to modify the protocol such that it can actually be realized, \cite{AG17} also present a variant in which all the values are given using polynomially many bits - resulting in inverse-exponential precision. In this variant, the completeness can no longer be maintained using exact equation checks. Instead, the protocol is adapted to use approximated comparisons $\approx_\mu$  where $\mu$ is inverse-exponential. As a by product, in this variant the verification is for the claim $\tr{A}\approx_\mu C$. \cite{AG17} show that the adapted protocol maintains a perfect completeness parameter, and has soundness $\frac{1}{3}$. To prove the soundness parameter for this imprecise protocol, \cite{AG17} consider the probability with which $P$ may decrease the error value $\tr{\Delta_i}$ throughout the protocol. They show that the probability with which a malicious prover may pass the protocol while decreasing a large $\tr{\Delta_0}$ throughout the protocol until making it small enough such that $\tr{\Delta_T}\approx_\mu0$ is at most $\frac{1}{3}$, and otherwise $V$ will reject in the final round.
	
\subsection{Cheating strategy against the AG protocol for $\BQP$ verification} \label{sec: AG cheat}
In light of the imprecise AG protocol, a natural question arises: what if the protocol is further adapted, to use approximated comparisons $\approx_\mu$ where $\mu$ is inverse-polynomial, rather than inverse-exponential? If such an adapted protocol is still sound and complete, this protocol could be used to verify $\BQP$ computations using a $\BQP$ prover! \cite{AG17} address this issue, and show that their protocol is no longer sound if such a relaxation is made. To show this, they present a strategy by which a cheating prover can convince the verifer to accept the false claim $\tr{A}=C$ for $A=\vert0^n\rangle \langle 0^n \vert G_T\cdot G_{T-1}\cdots G_1$ and value $C=\tr{A}+\frac{2}{3}$. In this strategy, the prover adaptively picks the error matrix $\Delta_i$, and chooses an appropriate matrix $m_i=M_i-\Delta_i$ (according to Equation \ref{eq:AG delta}) to send to the verifier. The strategy goes as follows\footnote{disregarding negligible imprecisions in order to only use polynomially many bits for each value.}:
	\begin{enumerate}
		\item {\em In the 0'th round} -- Denoting $\delta_0=\tr{A}-C$, $P$ picks $\Delta_0=\frac{\delta_0}{\tr{I}}I$, and sends $m_0=M_0-\Delta_0$ to $V$.
 		\item {\em In the $i$'th round} --  Denoting $\delta_i=\tr{\Delta_{i-1}\cdot g_i^{-1} \cdot u_i}$, $P$ picks $\Delta_i=\frac{\delta_i}{\tr{I}}I$, and sends $m_i=M_i-\Delta_i$ to $V$.
	\end{enumerate}
By the choice of $\delta_i$, $\Delta_i$ it is easy to see that 
\begin{equation} \label{eq:cheat consistency}
\tr{\Delta_i}=\frac{\delta_i}{\tr{I}}\tr{I}=\delta_i=\tr{\Delta_{i-1}\cdot g_i^{-1} \cdot u_i}
\end{equation}
\begin{equation} \label{eq:delta shrink}
\delta_i=\tr{\Delta_{i-1}\cdot g_i^{-1} \cdot u_i}=\delta_{i-1}\frac{\tr{g_i^{-1} \cdot u_i}}{\tr{I}}
\end{equation}

By Equations \ref{eq:delta equivalence}, \ref{eq:cheat consistency}, $P$ will manage to pass the first $T-1$ rounds. In addition, as $I$ has the maximal trace over all unitaries, Equation \ref{eq:delta shrink} means that $\delta_i\leq\delta_{i-1}$, and that there is some constant $\varepsilon < 1$ such tha $\delta_i \leq \varepsilon \cdot \delta_{i-1}$ for at least $\frac{T}{2}$ rounds\footnote{With probability (over the choices of the unitaries $u_1,\dots,u_T$) that approaches 1.}. This means, that with extremely high probability, $\delta_T$ is inverse-exponentially small, and so $V$ accepts after the final round as $\delta_T=\tr{\Delta_T}\approx_\mu 0$.

\section{Linear Scalar Consistency Checking protocols}\label{sec: LSCCs}

We would now like to define a family of protocols of a particular fairly natural structure, 
generalizing the AG protocol (as well as the protocol of  \cite{AAV13}). 
Recall that in The AG protocol the consistency checks are of the following form: multiply a given matrix $m_{i-1}$ passed from the prover to the verifier in the $\parent{i-1}$'th round by some other matrix $g_i^{-1}\cdot u_i$, calculate the trace, and compare it to the trace of a matrix $m_i$ which is sent by the prover on the $i$'th round. This is a special case of what we call "linear scalar consistency checks", where in order to compare two matrices we compare their values under some linear functional (the trace operator, in the AG protocol), up to some linear transformation on one of the matrices.
Consistency means that the two values are equal. 
With this in mind, we use $\Mk$ to denote the space of $k\times k$ matrices and define the following family of protocols, where at each round a consistency check is made using just a single, linearly calculated scalar:

\begin{defn} \textbf{(Linear-Scalar Consistency Checking)}\label{def:lscc protocols}

We say that a $T\parent{n}$-round interactive protocol between a verifier $\cV$ and a prover $\cP$ is a \textit{Linear-Scalar Consistency Checking (LSCC)} protocol for a scalar function $\cC:\{0,1\}^*\rightarrow \bC$ if there exists a linear scalar function $\cF: \Mk \rightarrow \bC$ such that given an input $x\in\{0,1\}^n$ and value $C\in\bC$:

\begin{enumerate}
    \item At round $0$: $\cV$ asks $\cP$ for some matrix $M_0\in \Mk$ s.t. $\cC\parent{x}=\cF\parent{M_0}$, receives a matrix $m_0$, and verifies that $C=\cF\parent{m_0}$ (rejects otherwise).
    \item At each round $0<i \leq T$: $\cV$ picks $\bT_i\sim\cD_i\parent{x}$ where $\cD_i$ is a distribution on the set of linear transformations on $\Mk$, with operator norm at most $1$, such that $\cD_i$ is computable by $\cV$. $\cV$ then sends 
    $\bT_i$ to $\cP$, and expects $\cP$ to send back a matrix $M_i \in \Mk$ s.t. $\cF\parent{\bT_i\parent{ M_{i-1}}}=\cF\parent{M_i}$. $\cV$ receives a matrix $m_i$, and checks for consistency by verifying $\cF\parent{\bT_i\parent{m_{i-1}}}=\cF\parent{m_i}$ (rejects otherwise). 
    \item After round $T$: $\cV$ accepts iff $\cF\parent{m_{T}}= f\parent{x, \bT_1,\dots,\bT_T} $, where $f$ is some function computable by $\cV$ (rejects otherwise).
\end{enumerate}
\end{defn}

To fully understand Definition \ref{def:lscc protocols}, consider the AG protocol. There, the input $x$ is a description of a quantum gate sequence $g_1,\cdots, g_T$, the function $\cC$ is the trace of the \textit{top row matrix} of the gate sequence (See Definition \ref{def:toprow}), $\cF$ is the trace operator, $\bT_i=g_i^{-1}\cdot u_i$, and $f=\tr{U_T\cdot U_{T-1}\cdots U_1 \cdot \vert0^n\rangle \langle 0^n \vert}$. \\

So we have a family of protocols which generalizes the AG protocol\footnote{as well as that of \cite{AAV13}, as all the consistency checks there consist of comparing the trace of a new matrix to the trace of a previous matrix, under some linear projection.}. However, in \textit{LSCC} protocols the consistency checks are exact; if we want to study protocols which are meant to be carried out with $\BQP$ provers it is necessary to allow approximations in consistency checks, 
since a $\BQP$ prover necessarily introduces inverse polynomial inaccuracies to its estimations of the matrix elements due to statistical errors. 
This suggests that we should consider protocols where such inverse polynomial inaccuracies in the verifier's consistency checks are allowed, unlike in the \textit{LSCC} family. 
We note that \cite{AG17} already allowed inaccuracies in their adapted (imprecise) protocol.
There, the inaccuracies were inverse-exponential. To allow inaccuracies in general \textit{LSCC} protocols, we define the following family called \textit{ILSCC}.  

\begin{defn}\label{def:ilscc protocols} \textbf{(Inexact Linear-Scalar Consistency Checking)} (Informal)

An \textit{Inexact Linear Scalar Consistency Checking (ILSCC)} protocol with precision $\mu\parent{n}$ for a scalar function $\cC:\{0,1\}^*\rightarrow \bC$ is defined similarly to an \textit{LSCC} protocol, except that in the $\ILSCC$ protocol all equality comparisons are done to within precision $\hat{\mu}$, where $\hat{\mu}=\mu\parent{n}\cdot \displaystyle\max_{y\in \{0,1\}^n}{\abs{\cC\parent{y}}}$. For a formal definition, see Appendix \ref{app: ILSCC protocols}.
\end{defn}

Having defined \textit{ILSCC} protocols, we would like to understand whether they can be used in interactive proofs for functional problems. That is, given an input, we would like to verify the value of some function on this input. In order to speak in a formal, general way about verification of such functional problems, we define a language $\cL$ based on the function. Given a function $\cC\parent{\cdot}$, the language $\cL$ contains all pairs of the form $\parent{x, \cC\parent{x}}$. However, similarly to previous notions we have presented - we would also like to allow a tolerance in our functional verification, so $\cL$ also contains pairs $\parent{x, C}$ where $C$ is an approximation of $\cC\parent{x}$. We pick an approximation that allows $\BQP$ verification (see Section \ref{sec: top row}).  Formally, we define the following:

\begin{defn} \textbf{(Scalar Function Language)}
For a given function $\cC:\{0,1\}^*\rightarrow\bC$, we define the set  $SF_n\parent{\cC}$ as $SF_n\parent{\cC}=\{\parent{x, C} \ | \ x\in\{0,1\}^n, \ \abs{\cC\parent{x} - C} \leq \frac{1}{6}\cdot\displaystyle\max_{y\in \{0,1\}^n}{\abs{\cC(y)}} \}$. 
We define the \it{Scalar Function Language} associated with the function $\cC$ as 
\[SF\parent{\cC}\equiv\cup_{n=0}^\infty SF_n\parent{\cC}.\]
\end{defn}

It will also be useful to define the class of all \textit{Scalar Functional Languages}: 

\begin{defn} \textbf{(Scalar Function Class)}We define the class SFL (for \textit{Scalar Function Languages}) 
to be the set of all scalar function languages, over all functions.
\end{defn}

Furthermore, for any language $\cL=SF\parent{\cC}$ we define a \textit{Polynomial Linear-Scalar Consistency Proof System}:
\begin{defn}\label{def:linear consistency proof system} \textbf{(Polynomial Linear-Scalar Consistency Proof System)}
Let $\cL=SF\parent{\cC}$ be a Scalar Functional Language associated with the function $\cC:\{0,1\}^*\rightarrow\bC$. 
 We say that $\cL$ has a \textit{Polynomial Linear-Scalar Consistency Proof System} if there exists a protocol $\cW$ between a verifier $\cV \in \BPP$ and a prover $\cP$ such that:
\begin{enumerate}
    \item $\cW$ is an \textit{ILSCC} protocol for $\cC$ with precision $\mu\parent{n}=\frac{1}{\poly{n}}$, 
    and $T(n)$ rounds such that $\Omega(n)\leq T(n) \leq \poly{n}$
    \item (Exact completeness) $\forall x\in \{0,1\}^n$, $\prob{\cW(\parent{x,\cC(x)}=Accept}=1$. Moreover, this property holds even if we set the precision $\mu$ of $\cW$ to be arbitrarily small
    \item (Soundness) For any protocol $\widetilde{\cW}$ between a verifier $\cV$ and a prover $\cP$, where $\cV$ behaves the same as in $\cW$ (but the prover may act adversarially): $\forall \parent{x, c} \notin\cL$, $\prob{\widetilde{\cW}\parent{x, c}=Accept}\leq\frac{1}{3}$ 
\end{enumerate}
\end{defn}
And we define the class of \textit{Polynomial Linear-Scalar Checkable} languages:
\begin{defn} \textbf{(Polynomial Linear-Scalar Checkable)}
The class \textit{Polynomial Linear-Scalar Checkable} ($PLSC$) is defined as the class of all Scalar Functional Languages which have a Polynomial Linear-Scalar Consistency Proof System. That is:
\[
PLSC=\{\cL \in SFL \ | \ \cL \text{ has a } \textit{Polynomial Linear-Scalar Consistency Proof System} \}
\]
\end{defn}
Finally, we define the following:
\begin{defn}\textbf{(Viable protocols)}
Let $\cL\in PLSC$.  
We call any protocol $\cW$ for which the three conditions of Definition \ref{def:linear consistency proof system} are met, a \textit{Viable protocol for} $\cL$. We denote the set of all Viable protocols for $\cL$ by $Viable\parent{\cL}$.
\end{defn}
\begin{defn}\textbf{(Dishonest protocols)}
Let $\cL\in PLSC$, and let $\cW\in Viable\parent{\cL}$. We call any protocol $\widetilde{\cW}$ where $\cV$ behaves the same as in $\cW$  (but the prover may act adversarially) a \textit{dishonest protocol for $\cL$ with respect to $\cW$}. We denote the set of all dishonest protocols for $\cL$ with respect to $\cW$ by $Dishonest\parent{\cL}_\cW$.
\end{defn}

\section{Limitations on $\ILSCC$ protocols for verifying $\BQP$} \label{sec: ILSCC limitations}
To gain insight on the class \textit{PLSC} of languages that are verifiable by \textit{ILSCC} protocols, we will prove two theorems concerning their possible viable protocols. 
To this end, we associate each $\ILSCC$ protocol with a parameter which we call the \textit{expected next value}. Intuitively, this parameter has to do with the maximal reduction of the scalar value from round to round: At round $i$ the verifier receives the matrix $m_i$, and evaluates its associated \textit{value} $\cF\parent{m_{i}}$. So for the matrix $m_i$, the \textit{next} value is $\cF\parent{\bT_{i+1}\parent{m_i}}$.
Now, let 
\[
\cQ= \displaystyle\argmax_{\norm{Q}_{F}=1}\parent{\abs{\cF(Q) }},\ q^* = \displaystyle\max_{\norm{Q}_{F}=1}\parent{\abs{\cF\parent{Q}}} 
\]
\\
Where we use $\norm{}_F$ to denote the Frobenius norm. We note that for any $Q\in\cQ$, the \textit{next} value which corresponds to $Q$ can not be greater than $Q$'s \textit{value} $q^*$, due to the fact that $\bT_{i+1}$ has operator norm at most 1. We define the \textit{expected next value} with respect to such $Q$, where the expectation is taken over the rounds and choices of random operators:
\begin{defn}{\bf (Expected next value)}
Let $\cL\in PLSC$, and let $\cW\in Viable\parent{\cL}$. We define $\cW$'s \textit{expected next value} on inputs in $\{\parent{x,C}$ \ | \ $x\in\{0,1\}^n,\ C\in\bC\}$, denoted $E_{\cW}^n$, to be the minimal expectation value 
\begin{equation}\label{eq:expected next value}
E_{\cW}^n = \displaystyle \min_{\substack{{x\in \{0,1\}^n}\\ Q \in\cQ}}{\parent{\E_{\substack{{\bT_i\sim\cD_i\parent{x}}\\0<i\leq T(n) }}\left[\abs{\cF\parent{\bT_i \parent{Q}}}\right]}}
\end{equation}

We denote one of the matrices at which the minimum is attained in Equation \ref{eq:expected next value} by $Q^*$. We will also use $E_\cW$ to refer to $E_\cW^n$ for an arbitrarily large $n$, in places where rigor may be sacrificed for simplicity.
\end{defn}

We note that $E_{\cW}^n\leq q^*$, and state our two theorems:
\begin{theorem*}\label{cl:decreasing max value}
$\forall \cL\in PLSC, \ \forall \cW\in Viable\parent{\cL}, \ \forall \varepsilon > 0, \ \exists N \ | \ \forall n > N  : \ E_\cW^n > \parent{1-\varepsilon}q^*$
\end{theorem*}

\begin{theorem*}\label{cl:non-decreasing max value}
Let $\cL\in PLSC$. If $\cL$ has a viable protocol $\cW\in Viable\parent{\cL}$ such that $\forall n\in \bN: \ E_\cW^n = q^*$, then $\cL\in\BPP$ 
\end{theorem*}

In other, less formal words, Theorem \ref{cl:decreasing max value} means that for any constant $\varepsilon>0$ there is no viable protocol $\cW$ with \textit{expected next value} $E_\cW \leq \parent{1-\varepsilon}q^*$. On the other hand, Theorem \ref{cl:non-decreasing max value} says that any language $\cL\in PLSC$ which has a viable protocol $\cW$ with $E_\cW=q^*$ is in $\BPP$, so in particular $\cL$ is not complete for $\BQP$ unless $\BQP\subseteq\BPP$.\\

The perceptive reader may have noticed that the previous Theorem \ref{cl:decreasing max value} and Theorem \ref{cl:non-decreasing max value} did not cover all possible values for the \textit{expected next value} $E_\cW$. Indeed, there remains a band of values for $E_\cW$ which is somewhat mysterious. We discuss this "band of mystery" in Section \ref{sec: open questions}.

\subsection{Theorem \ref{cl:decreasing max value} proof overview}\label{sec: decreasing value proof}
Given a language $\cL\in PLSC$ with an associated function $\cC:\{0,1\}^*\rightarrow\bC$ and $\cW\in Viable\parent{\cL}$, we would like to analyze the execution of any dishonest protocol $\widetilde{\cW}$ for $\cL$ with respect to $\cW$ on input $\parent{x, C}$ for some $x\in \{0,1\}^n$, $C\in\bC$. For this purpose, we define an error term $\Delta_i$ for each round of a specific execution of $\widetilde{\cW}$ on $\parent{x,C}$. To do so, we first define the honest matrix $M_i$ to be the matrix that is received from $\cP$ in the $i$'th round when running the (honest) viable protocol $\cW$ on input $(x, \cC\parent{x})$ with the same selection of $\bT_1,\dots,\bT_i$ by the verifier as in the run of $\widetilde{\cW}$. Keeping in mind that in our execution of $\widetilde{\cW}$ the prover sends a matrix $m_i$ in the $i$'th round, we define:
\[
\Delta_i=M_i-m_i
\]
By exact completeness, we have 
\begin{equation}
    \cF\parent{\bT_i \parent{ M_{i-1}}}=\cF\parent{M_i}
\end{equation}
So by the \rmarker{linearity} of $\cF$, we get that the consistency condition 
\begin{equation}
    \cF\parent{\bT_i\parent{m_{i-1}}}\approx_{\hat{\mu}} \cF\parent{m_i}
\end{equation}
is equivalent to the condition
\begin{equation}\label{eq:equivalent consistency check}
        \cF\parent{\bT_i\parent{\Delta_{i-1}}}\approx_{\hat{\mu}} \cF\parent{\Delta_i}
\end{equation}
\\

Now, to prove Theorem \ref{cl:decreasing max value}, we show that if there exists $\varepsilon > 0$ such that $E_\cW\leq\parent{1-\varepsilon}q^*$, then there exists a cheating strategy for the prover where $\abs{\cF\parent{\Delta_T}}$ is exponentially smaller than $\abs{\cF\parent{\Delta_0}}$. This strategy is realized by choosing values such that $\Delta_i=k_i\cdot Q^*$, where $k_i$ is a scalar which is iteratively picked as to pass the $i$'th round. Due to $Q^*$'s maximality under $\cF$, the \textit{error terms} $\abs{\cF\parent{\Delta_i}}=\abs{k_i\cdot q^*}$ are guaranteed to be monotonically decreasing. Moreover, since $E_\cW\leq\parent{1-\varepsilon}q^*$, there exists an input for which the error terms are likely to decrease by some constant factor in a constant fraction of the rounds - resulting with an exponential overall decrease. This is a generalization of the cheating strategy against the AG protocol (see Section \ref{sec: AG cheat}), where $Q^*$ is just the (normalized) identity matrix $I$. For a detailed proof, see Appendix \ref{app: decreasing max value proof}.

\subsection{Theorem \ref{cl:non-decreasing max value} proof}\label{sec:non-decreasing max value proof}
\begin{proof}

Let $\cL\in PLSC$, $\cW\in Viable\parent{\cL}$ with respective operator $\cF$ and distributions $\{\cD_i\}_{i=1}^{T\parent{n}}$ such that $\forall n\in\bN : E_\cW^n = E_\cW = q^*$. In order to show that $\cL\in\BPP$ we first show that for any $A\in \Mk$, its value under the operation of $\cF$ isn't changed by first operating on it with $\bT_i\sim\cD_i$ (up to a phase change which doesn't depend on $A$). More formally, we claim the following:

\begin{claim}\label{cl:stabilization}
Let $\cL\in PLSC$, $\cW\in Viable\parent{\cL}$ with respective operator $\cF$ and distributions $\{\cD_i\}_{i=1}^{T}$, and let $A\in \Mk$. If $E_\cW = q^*$ then for all $0<i\leq T$:
\[
    \displaystyle\Pr_{\bT_i\sim\cD_i}[\cF(A) = \lambda\parent{\bT_i}\cdot\cF(\bT_i\parent{A})] = 1
\]
Where $\lambda\parent{\bT_i}$ is a scalar that only depends on $\bT_i$, such that $\abs{\lambda}=1$.
\end{claim}
 Claim \ref{cl:stabilization} is proven in Appendix \ref{app: stabilization proof}. \\

Now we explain why Claim \ref{cl:stabilization} entails $\cL\in\BPP$. Let us consider any accepting run of the protocol $\cW$ (with a choice of $\bT_1, \dots, \bT_T$ and $m_0,\dots,m_T$ by the verifier $\cV$ and prover $\cP$ respectively). By recursively applying Claim \ref{cl:stabilization} we get that, up to phase changes due to $\lambda\parent{\bT_1},\dots,\lambda\parent{\bT_T}$:
\[
\cF\parent{m_0}=\cdots=\cF\parent{m_T}
\]
So $\cP$ could pass the first $T-1$ rounds by simply sending $m_0$ over and over again\footnote{W.l.o.g., $\cV$ can be assumed to make the phase changes on its own, as $\lambda\parent{\bT_i}$ only depends on $\bT_i$.}. Then, to pass the $T$'th round, $\cP$ can simply send $m_T$. This makes it clear that the intermediate rounds of interaction are in fact completely redundant. The protocol is exactly equivalent to a 2 round protocol $\cW_2$ where
\begin{itemize}
    \item in the first round: $\cP$ sends $m_0$ to $\cV$
    \item in the second round: $\cV$ sends $\bT_1, \dots, \bT_T$ to $\cP$, and $\cP$ sends back $m_T$ ($\cV$ then performs its verification on $m_T$ using $\bT_1, \dots, \bT_T$)  
\end{itemize}
This equivalence means that $\cW_2$ has the same soundness and completeness properties as $\cW$, which puts $\cL$ is in the complexity class $\MA{3}$ (Merlin Arthur Merlin) \cite{Bab85}. Now, if we take into consideration that the equality checking of $\cV$ is only done with inverse polynomial precision, it follows that $\cP$ can also pass the protocol using the inverse-polynomial approximations $m_0, m_T$ which are obtained from $m_0, m_T$ by only keeping logarithmically many bits to describe each. This means that there are only polynomially many possible choices of $m_0$, and for each such choice - given also a choice of $\bT_1, \dots, \bT_T$ - there are only polynomially many possible options for $m_T$. This means that $\cV$ can simulate $\cP$ on her own, which results in $\cL\in\BPP$ as required.
\end{proof}

We note that the final argument in the proof of Theorem \ref{cl:non-decreasing max value} relied on an implicit assumption that the matrices are in $\Mk$ where $k$ is constant. But one could also consider protocols where matrix sizes are polynomial - for instance, if they represent a reduced density matrix on a logarithmic number of qubits. This change will imply that the verifier could no longer simulate the protocol on her own, resulting in $\cL \in \MA{3}=\AM$ \cite{BM88} instead of $\cL\in\BPP$. This change would mean that in particular $\cL$ is not complete for $\BQP$ unless $\BQP\subseteq\AM$, which is weaker than the original corollary but still seems rather unlikely.

\section{Discussion and open questions}\label{sec: open questions}
We hope our definition and analysis of $\ILSCC$s help shed light on the innate difficulties of using natural approaches for $\BQP$ verification. To this end, we remark that our discussion of $\ILSCC$ protocols might be more general than it appears in first glance. 
\begin{enumerate}
    \item We had already mentioned in Section \ref{sec: intro} that protocols where consistency checks are linear are natural to consider with respect to $\BQP$ problems, as quantum computers are considered to be linear beings. However, one could consider using vectors for consistency checking, rather than scalars. We note that this only has the potential to circumvent Theorem \ref{cl:decreasing max value} if the consistency checking involved \textit{non-linear} operations on the vectors (e.g. a consistency check which involves norms). Otherwise, there is an equivalent protocol where the consistency test only involves scalars.
    \item  We would also like to address the fact that $\ILSCC$ protocols use a \textit{fixed} functional $\cF$ for consistency checking. Indeed, the cheating strategy (Section \ref{sec: decreasing value proof}; Appendix \ref{app: decreasing max value proof}) uses the prover's knowledge of $\cF$ to pick the error matrices $\Delta_i$ according to $Q^*$. To circumvent this, one could consider a protocol where the consistency tests are different for each round. So in the $i$'th round the consistency test will be of the form $\hat{\cF}_i(m_{i-1})=\widetilde\cF_i(m_i)$ for $\hat{\cF}_i,\ \widetilde\cF_i$ which are distinct random linear functionals which are chosen separately for each round\footnote{\textit{ILSCC} protocols can be generalized to include such protocols by introduction another random linear transformation $\widetilde{\bT}_i$ for the i'th round and defining $\widetilde\cF_i=\cF\circ\widetilde{\bT}_i; \ \hat{\cF}_i=\widetilde{\cF}_{i-1}\circ\bT_i$, and using consistency checks of the form $\hat{\cF}_i(m_{i-1})=\widetilde\cF_i(m_i)$.}. The caveat is that the prover still needs to be told which matrix $m_i$ he should send in order to pass the $i$'th round, and the prover could use this knowledge to infer the functional $\widetilde{\cF}_i$ - resulting with the same problem.
\end{enumerate}

That being said, our analysis does bring forth open questions regarding the possibility of $\BQP$ verification, even without finding non-linear structure in a $\BQP$ complete problem. We address a couple of them in particular:
\begin{enumerate}
\item \textbf{Number of rounds vs. precision} \quad The cheating strategy presented in Section \ref{sec: decreasing value proof} (and Appendix \ref{app: decreasing max value proof}) is based on repeatedly decreasing an initial error term such that by the last round it is smaller than an inaccuracy parameter $\hat{\mu}$ which is needed for the protocol's completeness. This suggests another approach to mitigate the cheating strategy: one could look for protocols where the number of rounds is such that even if at each round the error term is decreased by e.g. a constant factor - it is still not smaller than the inaccuracy value. To this end one could try two venues:
\begin{itemize}
    \item \textbf{Decrease the number of rounds} - One could try to find a protocol which verifies a $\BQP$-complete problem and has a sufficiently sub-linear number of rounds that a final error term is inverse-polynomial even if an initial error term decreases by a constant factor at each round with high probability. We note that such a result will put $\BQP$ in a subclass class of $\IP$ which will be an interesting result in its own merits.
    \item \textbf{Decrease the allowed inaccuracy} - On the other hand, one could look for a protocol where the prover is not allowed inverse-polynomial inaccuracy to begin with. The motivation behind allowing the prover such inaccuracies was that a $\BQP$ machine experiences such inaccuracies in its measurements. But what if the prover could be expected to compute matrix entries with higher precision? For example, if there were only a polynomial number of \textit{discrete} possible values\footnote{Assuming there are still enough possible values, so the protocol's soundness isn't hurt due to insufficient randomness.}, which are all inverse-polynomially separated from each-other, then the prover can be expected to infer the \textit{exact} values with high probability by inverse-polynomially approximating them. 
\end{itemize}
\item \textbf{\textit{ILSCC} band of mystery} \quad As previously mentioned, the analysis of Section \ref{sec: ILSCC limitations} did not cover all possible \textit{expected next values}. One could consider a protocols $\cW$ where $E_\cW = (1-\varepsilon) q^*$ where $\varepsilon$ is not bounded away from 0 by a constant. We first mention that indeed, our analysis was not tight. 
Instead of Claim \ref{cl:stabilization}, it is easy to prove the following inexact variant:
\begin{claim}
For any $A\in \Mk$, if $E_\cW \ge \left(1-\frac{1}{f(n)}\right)q^*$ for a super-polynomial $f(n)$ then there exist super-polynomial functions $\hat{f}(n)$, $\tilde{f}(n)$ such that if we denote $\alpha=\frac{1}{\hat{f}(n)}$, $\beta=\frac{1}{\tilde{f}(n)}$ 
\begin{equation}\label{eq:inexact general stability}
    \displaystyle\Pr_{\bT_i\sim\cD_i}\left[\cF(A) \approx_\alpha \cF(\bT_i\parent{A})\right] \approx_\beta 1
\end{equation}
\end{claim}

Using the union bound, Equation (\ref{eq:inexact general stability}) can then be used to extend Theorem \ref{cl:non-decreasing max value} to exclude the region $E_\cW \ge \left(1-\frac{1}{f(n)}\right)q^*$. \\
In the other direction, Theorem \ref{cl:decreasing max value} can be extended to show $E_\cW > \left(1-\frac{1}{g(T)}\right)q^*$ for some sub-linear function $g(T)$, which excludes the region where $E_\cW \leq \left(1-\frac{1}{g(T)}\right)q^*$. But still, there remains a band which is not covered by the two cases. This is not a coincidence. To demonstrate the band where neither of the cases apply, let us consider a hypothetical example where $T(n)=n$, with a choice of $\cD_i$'s such that
\begin{equation}
    \displaystyle\Pr_{\bT_i\sim\cD_i}\left[\cF(\bT_i\parent{Q^*}) = \left(1-\frac{1}{n}\right)\cF(Q^*) \right] = 1 
\end{equation}
In this scenario, the \textit{cheating strategy} and arguments of Theorem \ref{cl:decreasing max value} only imply that $\cF\parent{\Delta_T}$ is smaller than $\cF\parent{\Delta_0}$ by a constant factor. So the prover can not decrease his error enough as to pass the verifier's final verification, and his proposed cheating strategy is not valid. 
On the other hand, in this example the verifier's consistency tests can be accurate enough to pick up on the change of the values $\cF(m_i)$ of the intermediate $m_i$ matrices. This means the intermediate rounds can no longer be skipped - and the arguments of Theorem \ref{cl:non-decreasing max value} are no longer valid. 
\end{enumerate}


\paragraph{Acknowledgments}
The author acknowledges the generous support of ERC grant number 280157.
Also, many thanks to Dorit Aharonov for useful discussions both during and prior to this work, and to Zuzana Gavorova for useful discussions.

\bibliographystyle{alpha}
\bibliography{InformationTheoreticVerification.bib}

\newcommand{\etalchar}[1]{$^{#1}$}
\begin{thebibliography}{ABOEM17}

\bibitem[Aar05]{Aar05}
Scott Aaronson.
\newblock Quantum computing, postselection, and probabilistic polynomial-time.
\newblock In {\em Proceedings of the Royal Society of London A: Mathematical,
  Physical and Engineering Sciences}, volume 461, pages 3473--3482. The Royal
  Society, 2005.

\bibitem[AAV13]{AAV13}
Dorit Aharonov, Itai Arad, and Thomas Vidick.
\newblock Guest column: the quantum pcp conjecture.
\newblock {\em Acm sigact news}, 44(2):47--79, 2013.

\bibitem[ABOE10]{ABOE10}
Dorit Aharonov, Michael Ben-Or, and Elad Eban.
\newblock Interactive proofs for quantum computations.
\newblock {\em arXiv version: arXiv:0810.5375 (2008), conference version:
  Proceedings of the 2010 Conference on Innovations in Theoretical Computer
  Science}, 2010.

\bibitem[ABOEM17]{ABOEM17}
Dorit Aharonov, Michael Ben-Or, Elad Eban, and Urmila Mahadev.
\newblock Interactive proofs for quantum computations.
\newblock {\em arXiv preprint arXiv:1704.04487}, 2017.

\bibitem[AG17]{AG17}
Dorit Aharonov and Ayal Green.
\newblock A quantum inspired proof of $\p^{\sharpP} \subseteq \mathsf{IP}$.
\newblock {\em arXiv preprint arXiv:1710.09078}, 2017.

\bibitem[Bab85]{Bab85}
L{\'a}szl{\'o} Babai.
\newblock Trading group theory for randomness.
\newblock In {\em Proceedings of the seventeenth annual ACM symposium on Theory
  of computing}, pages 421--429, 1985.

\bibitem[BFK09]{BFK09}
Anne Broadbent, Joseph Fitzsimons, and Elham Kashefi.
\newblock Universal blind quantum computation.
\newblock In {\em 2009 50th Annual IEEE Symposium on Foundations of Computer
  Science}, pages 517--526. IEEE, 2009.

\bibitem[BH13]{BH13}
Fernando~GSL Brandao and Aram~W Harrow.
\newblock Product-state approximations to quantum ground states.
\newblock In {\em Proceedings of the forty-fifth annual ACM symposium on Theory
  of computing}, pages 871--880. ACM, 2013.

\bibitem[BM88]{BM88}
L{\'a}szl{\'o} Babai and Shlomo Moran.
\newblock Arthur-merlin games: a randomized proof system, and a hierarchy of
  complexity classes.
\newblock {\em Journal of Computer and System Sciences}, 36(2):254--276, 1988.

\bibitem[BV97]{BV93}
Ethan Bernstein and Umesh Vazirani.
\newblock Quantum complexity theory.
\newblock {\em SIAM Journal on Computing}, 26(5):1411--1473, 1997.

\bibitem[CGJV19]{CGJV19}
Andrea Coladangelo, Alex~B Grilo, Stacey Jeffery, and Thomas Vidick.
\newblock Verifier-on-a-leash: New schemes for verifiable delegated quantum
  computation, with quasilinear resources.
\newblock In {\em Annual International Conference on the Theory and
  Applications of Cryptographic Techniques}, pages 247--277. Springer, 2019.

\bibitem[FH13]{FH13}
Michael~H Freedman and Matthew~B Hastings.
\newblock Quantum systems on non-$ k $-hyperfinite complexes: A generalization
  of classical statistical mechanics on expander graphs.
\newblock {\em arXiv preprint arXiv:1301.1363}, 2013.

\bibitem[FK17]{FK17}
Joseph~F Fitzsimons and Elham Kashefi.
\newblock Unconditionally verifiable blind quantum computation.
\newblock {\em Physical Review A}, 96(1):012303, 2017.

\bibitem[FV15]{FV15}
Joseph Fitzsimons and Thomas Vidick.
\newblock A multiprover interactive proof system for the local hamiltonian
  problem.
\newblock In {\em Proceedings of the 2015 Conference on Innovations in
  Theoretical Computer Science}, pages 103--112. ACM, 2015.

\bibitem[Ji16]{Ji16a}
Zhengfeng Ji.
\newblock Classical verification of quantum proofs.
\newblock In {\em Proceedings of the forty-eighth annual ACM symposium on
  Theory of Computing}, pages 885--898. ACM, 2016.

\bibitem[Ji17]{Ji17}
Zhengfeng Ji.
\newblock Compression of quantum multi-prover interactive proofs.
\newblock In {\em Proceedings of the 49th Annual ACM SIGACT Symposium on Theory
  of Computing}, pages 289--302, 2017.

\bibitem[JNV{\etalchar{+}}20]{JNVWY20}
Zhengfeng Ji, Anand Natarajan, Thomas Vidick, John Wright, and Henry Yuen.
\newblock Quantum soundness of the classical low individual degree test.
\newblock {\em arXiv preprint arXiv:2009.12982}, 2020.

\bibitem[LFKN92]{LFKN92}
Carsten Lund, Lance Fortnow, Howard Karloff, and Noam Nisan.
\newblock Algebraic methods for interactive proof systems.
\newblock {\em Journal of the ACM (JACM)}, 39(4):859--868, 1992.

\bibitem[Mah18]{Mah18}
Urmila Mahadev.
\newblock Classical verification of quantum computations.
\newblock In {\em 2018 IEEE 59th Annual Symposium on Foundations of Computer
  Science (FOCS)}, pages 259--267. IEEE, 2018.

\bibitem[MF16]{FM16}
Tomoyuki Morimae and Joseph~F Fitzsimons.
\newblock Post hoc verification with a single prover.
\newblock {\em arXiv preprint arXiv:1603.06046}, 2016.

\bibitem[NV17]{NV17}
Anand Natarajan and Thomas Vidick.
\newblock A quantum linearity test for robustly verifying entanglement.
\newblock In {\em Proceedings of the 49th Annual ACM SIGACT Symposium on Theory
  of Computing}, pages 1003--1015. ACM, 2017.

\bibitem[NV18]{NV18}
Anand Natarajan and Thomas Vidick.
\newblock Low-degree testing for quantum states, and a quantum entangled games
  pcp for qma.
\newblock In {\em 2018 IEEE 59th Annual Symposium on Foundations of Computer
  Science (FOCS)}, pages 731--742. IEEE, 2018.

\bibitem[RUV13]{RUV13}
Ben~W Reichardt, Falk Unger, and Umesh Vazirani.
\newblock Classical command of quantum systems.
\newblock {\em Nature}, 496(7446):456, 2013.

\bibitem[Sha92]{Sha92}
Adi Shamir.
\newblock Ip= pspace.
\newblock {\em Journal of the ACM (JACM)}, 39(4):869--877, 1992.

\end{thebibliography}

\appendix
\clearpage
\section{Theorem \ref{cl:decreasing max value} proof} \label{app: decreasing max value proof}
\paragraph{Theorem \ref{cl:decreasing max value}.}
$\forall \cL\in PLSC, \ \forall \cW\in Viable\parent{\cL}, \ \forall \varepsilon > 0, \ \exists N \ | \ \forall n > N  : \ E_\cW^n > \parent{1-\varepsilon}q^*$
\begin{proof}
Let $\cL\in PLSC$ be a language with viable protocol $\cW\in Viable\parent{\cL}$ such that $E_\cW^n=\parent{1-\varepsilon}q^*$ for some $\varepsilon > 0$, $n\in \bN$, and let $x\in \{0,1\}^n, \ C\in\bC$. We define the following \textit{cheating strategy}\footnote{throughout this analysis, we disregard inverse-exponential inaccuracies that occur as all communication in the protocol must be polynomial, as these inaccuracies are negligible with respect to the inverse-polynomial inaccuracies allowed in the comparisons.} for the prover $\cP$ on input $\parent{x, C}$ by recursively defining for each round an \textit{error value} $\delta_i$ along with a choice of \textit{error term} $\Delta_i$ to satisfy this \textit{error value}, which in turn commands the choice of matrices $m_i$ that the prover sends in each round (according to the definition of $\Delta_i$ in Section \ref{sec: decreasing value proof}):
\begin{enumerate}
    \item At round 0 we define $\delta_0=\cC\parent{x}-C$, $\Delta_0=\frac{C_0}{q^*}Q^*$
    \item At each round  $0<i \leq T$: we define $\delta_i=\cF\parent{\bT_i\parent{\Delta_{i-1}}}$, $\Delta_i=\frac{\delta_i}{q^*}Q^*$
\end{enumerate}
We now use the \textit{cheating strategy} to show that if $n$ is sufficiently large, $\cW$ is not sound. This means it can't be a \textit{viable protocol}, thus proving the theorem. To do so, let $C=\cC\parent{x}+\frac{1}{2}\displaystyle\max_{y\in \{0,1\}^n}{\abs{\cC\parent{y}}}$. We look at the probability that the \textit{cheating strategy} will make $\cV$ accept the input $\parent{x, C}\notin \cL$:
\begin{enumerate}
    \item At round 0: $\cV$ doesn't reject, as by linearity of $\cF$ we have \begin{equation}\cF\parent{\Delta_0}=\frac{\delta_0}{q^*}\cF\parent{Q^*}=\delta_0\end{equation}
and so
\begin{equation}
C=\cC\parent{\cI}-\delta_0=\cF\parent{M_0}-\cF\parent{\Delta_0}=\cF\parent{M_0-M_0+m_0}=\cF\parent{m_0}
\end{equation}

where we used perfect completeness for the second equality, and $\cF$'s linearity for the third equation.
    \item At each round $\cV$ doesn't reject, as the condition of Equation \ref{eq:equivalent consistency check} is met:
    \begin{equation}
    \cF\parent{\bT_i\parent{\Delta_{i-1}}}=\delta_i=\frac{\delta_i}{q^*}\cF\parent{Q^*}=\cF\parent{\Delta_i}
    \end{equation}
    \item Now we just need to show that $\cV$ accepts after round $T$ w.p. greater than $\frac{1}{3}$. To do so, we note that by perfect completeness:
    \begin{equation}
    f\parent{x, \bT_1,\dots,\bT_T}=\cF\parent{M_{T}}=\cF\parent{M_T-m_T+m_T}=\cF\parent{\Delta_T+m_T}=\cF\parent{m_T}+\delta_T
    \end{equation}
    So, keeping in mind that $\abs{\delta_0}=\frac{1}{2}\displaystyle\max_{y\in \{0,1\}^n}{\abs{\cC\parent{y}}}$, it suffices to show that 
    \begin{equation}\label{eq:probability of decrease}
        \prob{\abs{\delta_T}<\frac{\abs{\delta_0}}{poly\parent{n}}}>\frac{1}{3}
    \end{equation}
    Which would mean that $\prob{\cV \ Accepts}>\frac{1}{3}$ in the dishonest protocol where $\cV$ acts according to $\cW$ and the prover acts according to the cheating strategy, making $\cW$ not sound. For each round $0 < i \leq T$, we define the \textit{shrinkage} $S_i=\abs{\frac{\delta_i}{\delta_{i-1}}}$, and note that by linearity
    \begin{equation}\label{eq:round shrinkage}
   \delta_i=\frac{\delta_{i-1}}{q^*}\cF\parent{\bT_i\parent{Q^*}}\Rightarrow S_i = \frac{\abs{\cF\parent{\bT_i\parent{Q^*}}}}{q^*}
    \end{equation}
    We recall that $E_\cW^n=\parent{1-\varepsilon}q^*$, so there are at least $\frac{\varepsilon\cdot T}{2}$ rounds $i$ for which $\displaystyle\bE_{T_i}\left[S_i\right]\leq1-\frac{\varepsilon}{2}$. Let us denote the set of such rounds by $\S$. Similarly, $\forall i\in \S \ : \ \prob{S_i \le 1-\frac{\varepsilon}{4}}>\frac{\varepsilon}{4}$. Now we can denote $\xi_i$ the indicator variable for the event that $S_i \leq 1-\frac{\varepsilon}{4}$, and denote $\chi=\{i\in\S \ s.t \ \xi_i=1 \}$. A simple chernoff bound now gets us that $\prob{\abs{\chi}<\frac{1}{2}\cdot\frac{\varepsilon\cdot T}{2}\cdot\frac{\varepsilon}{4}}$ is inverse-exponentially small. 
    
    It remains to note that by the maximality of $q^*$ in equation \ref{eq:round shrinkage}, $S_i\leq 1$ for every round $0<i\leq T$. So we have $\abs{\delta_T} \leq \frac{\abs{\delta_0}}{\parent{1-\frac{\varepsilon}{4}}^{\abs{\chi}}}$. This means that inequality \ref{eq:probability of decrease} holds, as we have shown that the probability of $\abs{\chi}$ to be linear in $T$ is inverse-exponentially close to 1.
\end{enumerate}

\end{proof}

\clearpage
\section{Claim \ref{cl:stabilization} proof} \label{app: stabilization proof}
\paragraph{Claim \ref{cl:stabilization}.}
Let $\cL\in PLSC$, $\cW\in Viable\parent{\cL}$ with respective operator $\cF$ and distributions $\{\cD_i\}_{i=1}^{T}$, and let $A\in \Mk$. If $E_\cW = q^*$ then for all $0<i\leq T$:
\begin{equation}\label{eq:general stability}
       \displaystyle\Pr_{\bT_i\sim\cD_i}[\cF(A) = \lambda\parent{\bT_i}\cdot\cF(\bT_i\parent{A})] = 1
\end{equation}
Where $\lambda\parent{\bT_i}$ is a scalar that only depends on $\bT_i$, such that $\abs{\lambda}=1$.

\begin{proof}
To prove Claim \ref{cl:stabilization}, we first note that by the condition $E_\cW = q^*$ we get that:
\begin{equation}\label{eq:special stability}
    \displaystyle\Pr_{\bT_i\sim\cD_i}[\abs{\cF(Q^*)} = \abs{\cF(\bT_i\parent{Q^*})}] = 1
\end{equation}
But why does (\ref{eq:special stability}) imply (\ref{eq:general stability})? Why can't it be he case that $\bT_i$ doesn't change the value of $Q^*$, but decreases the values of other matrices? To understand this question better and answer it, we use the linear nature of $\bT_i$ and $\cF$. To this end, one might find it helpful to think of $Q^*$ and $A$ as vectors of dimension $k^2$, and of $\bT_i$ as a $k^2\times k^2$ matrix. in these terms, $\cF$'s \rmarker{linearity} means that it is simply an inner product with a vector $F$ of the same dimension\footnote{$\cF$ is a linear function on a $k^2$ dimension space, so it is a matrix with $k^2$ columns. and it is scalar, so it has a single row.} (e.g. the vector which corresponds to the trace operator is just the vector representation of the Identity matrix), whereas $\bT_i\parent{A}=\bT_i\cdot A$. Let us denote $F$'s normalized vector by $e:=\frac{F}{\norm{F}}$, so we have 
\begin{equation}
    \norm{e}=1, \  F=\norm{F}\cdot e
\end{equation}
It follows that, up to a phase change which we can omit w.l.o.g.:
\begin{equation}\label{eq:linear representation}
    Q^*=e, \ q^*=\norm{F} ,\ \cF(x) = q^*\innerprod{e}{x}
\end{equation}
Now, let us consider a basis $\{\Tilde{e}\}$ for $\Mk$ where $e$ is the first basis vector $e_1$, and denote $B$ the base change matrix, so that $B \cdot e=e_1$. In this basis we map $\bT_i$ to the operator $\cT$ by conjugating it by $B$:
\begin{equation} \label{eq:cT}
    \cT = B \circ \bT_i \circ B^\dagger
\end{equation}
From Equations (\ref{eq:special stability}), (\ref{eq:linear representation}) and (\ref{eq:cT}) we get:
\begin{equation}\label{eq:re special stability abs}
        \displaystyle\Pr_{\cT}[\abs{\innerprod{e_1}{\cT e_1}}=1] = 1
\end{equation}
Given the fact that $\cT$ has operator norm at most 1, Equation \ref{eq:re special stability abs} means that $e_1$ is an eigen vector of $\cT$ with eigen value $\lambda_1$ such that $\abs{\lambda_1}=1$. So we can rewrite (\ref{eq:re special stability abs}) as
\begin{equation}\label{eq:re special stability}
        \displaystyle\Pr_{\cT}[\lambda_1^\dagger\innerprod{e_1}{\cT e_1}=1] = 1
\end{equation}
Now we are ready to rewrite (\ref{eq:general stability}) as
\begin{equation} \label{eq:re general stability}
     \displaystyle\Pr_{\cT}[\innerprod{e_1}{A}=\lambda_1^\dagger\innerprod{e_1}{\cT\cdot A}] = 1
\end{equation}
And ask the (restated) question of whether (\ref{eq:re special stability}) implies (\ref{eq:re general stability}) in a more concrete way: why can't it be the case that $\cT$ is chosen such that it multiplies $e_1$ by $\lambda_1$, but acts on the rest of the space in a very random way? We use the fact that $\cT$ is also a \bmarker{linear} transformation on $\Mk$, and hence can be represented by a $k^2 \times k^2$ matrix. The answer is quite simple when considering the properties of this matrix. By (\ref{eq:re special stability}) we get that $\cT$'s top left entry has to be $\lambda_1$, as it sends $e_1$ to $\lambda_1\cdot e_1$. This means that the rest of the top row entries are 0: otherwise, if there is a column $i\neq 1$ where the top row has an entry $\varepsilon \neq 0$, we get for $v=\frac{1}{\sqrt{1+\varepsilon^2}}\left(e_1+\varepsilon\cdot e_i \right)$ that $\norm{v}=1$ but $\norm{\cT \cdot v} = \sqrt{1+\varepsilon^2}>1$, in contradictions to $\cT$ having a \bmarker{maximal eigen value} of 1. So if we denote the subspace which is orthogonal to $e_1$ by $e_1^\perp$, we get that while $\cT$ might indeed act in a non-trivial way on any \textit{vector} $v_\perp \in e_1^\perp$, it will not change its \textit{projection on $e_1$}. It is now easy to see that for a general vector $v\in\Mk$, if we decompose it as $v=c\cdot e_1 + v_\perp$ (for some $v_\perp\in e_1^\perp$ and scalar $c$) then:
\begin{equation}
    \lambda_1^\dagger\innerprod{e_1}{\cT v}=
    \lambda_1^\dagger\innerprod{e_1}{c\cdot \cT e_1 + \cT v_\perp} = c = \innerprod{e_1}{c\cdot e_1 + v_\perp} =     \innerprod{e_1}{v}
\end{equation}
as required, and we have proven Claim \ref{cl:stabilization}.
\end{proof}

\clearpage
\section{$\ILSCC$ protocols formal definition} \label{app: ILSCC protocols}
\paragraph{Definition \ref{def:ilscc protocols} (Linear-Scalar Consistency Checking)} (formal)

We say that a $T\parent{n}$-round interactive protocol between a verifier $\cV$ and a prover $\cP$ is an \textit{Inexact Linear Scalar Consistency Checking (ILSCC)} protocol with precision $\mu\parent{n}$ for a scalar function $\cC:\{0,1\}^*\rightarrow \bC$ if there exists a linear scalar function $\cF: \Mk \rightarrow \bC$ such that given an input $x\in\{0,1\}^n$ and $C\in\bC$:

\begin{enumerate}
    \item The maximal possible value of $\abs{\cC\parent{\cdot}}$ over $\{0,1\}^n$ is $\cC_{\max}$, and we denote $\hat{\mu} = \mu\parent{n} \cdot \cC_{\max}$
    \item At round $0$: $\cV$ asks $\cP$ for some matrix $M_0\in \Mk$ s.t. $\cC=\cF\parent{M_0}$, receives a matrix $m_0$, and verifies that $C\approx_{\hat{\mu}}\cF\parent{m_0}$ (rejects otherwise).
    \item At each round $0<i \leq T$: $\cV$ picks $\bT_i\sim\cD_i\parent{x}$ where $\cD_i$ is a distribution on the set of linear transformations on $\Mk$ with operator norm at most $1$, such that $\cD_i$ is computable by $\cV$. $\cV$ then asks $\cP$ for a matrix $M_i \in \Mk$ s.t. $\cF\parent{\bT_i\parent{ M_{i-1}}}=\cF\parent{M_i}$, receives a matrix $m_i$, and checks for consistency by verifying $\cF\parent{\bT_i\parent{m_{i-1}}}\approx_{\hat{\mu}}\cF\parent{m_i}$ (rejects otherwise)
    \item After round $T$: $\cV$ accepts iff $\cF\parent{m_{T}}\approx_{\hat{\mu}} f\parent{x, \bT_1,\dots,\bT_T} $, where $f$ is some function computable by $\cV$ (rejects otherwise).
\end{enumerate}

\end{document}